\documentclass[prd,aps,amsfonts,showpacs,longbibliography,notitlepage,twocolumn,superscriptaddress,nofootinbib]{revtex4-1}
\usepackage[dvipsnames]{xcolor}
\usepackage{multirow}
\usepackage{tabularx}
\usepackage{amsmath}
\usepackage{graphicx}

\usepackage[colorlinks=true, urlcolor=violet, linkcolor=blue, citecolor=red, hyperindex=true, linktocpage=true]{hyperref}

\usepackage{slashed}

\allowdisplaybreaks

\numberwithin{thm}{section}

\makeatletter
\renewcommand{\p@subsection}{}
\renewcommand{\p@subsubsection}{}
\makeatother

\usepackage{xcolor}
\usepackage{mathtools}
\usepackage{subcaption}
\usepackage[font=footnotesize]{caption}
\usepackage{ragged2e}

\newcommand{\figref}[1]{Fig.\;\ref{#1}}

\newcommand{\appref}[1]{App.\;\ref{#1}}

\newcommand{\ud}{\mathrm{d}}

\newcommand\cT{\mathcal{T}}

\newcommand\cP{\mathcal{P}}

\newcommand\cL{\mathcal{L}}

\newcommand{\ii}{\mathrm{i}}

\newcommand{\vect}[1]{\boldsymbol{#1}}
\newcommand{\im}{\mathrm{Im}~}
\newcommand{\re}{\mathrm{Re}~}

\newcommand{\p}{\partial}

\usepackage{dsfont}

\begin{document}

\title{Breakdown of hydrodynamics in a Galilean quantum Hall crystal }

\author{Xiaoyang Huang}
 \email{xiaoyang.huang@colorado.edu}
\affiliation{Department of Physics and Center for Theory of Quantum Matter, University of Colorado, Boulder, CO 80309, USA}

\author{Andrew Lucas}
\affiliation{Department of Physics and Center for Theory of Quantum Matter, University of Colorado, Boulder, CO 80309, USA}

\date{\today}

\begin{abstract}
We construct a nonlinear fluctuating hydrodynamic effective field theory for Galilean-invariant quantum Hall systems with spontaneously broken translational symmetry. 
Neglecting the role of energy conservation in a low-temperature regime, the hydrodynamic mode is a magnetophonon with quartic attenuation: $\omega\sim \pm k^2-\ii k^z$ with $z=4$. 
However, this linear response theory is unstable, and flows to a non-trivial dynamical universality class with $z\approx 3$.  We observe this scaling in 
numerical simulations of many-body classical Hamiltonian dynamics, in a model of an electronic crystal in the lowest Landau level.  Observing this magnetophonon decay rate in a quantum Hall crystal represents a promising setting to detect an analogue of a ``fractonic dynamical universality class" in a solid-state system, e.g. using microwave impedance microscopy.

\end{abstract}

\date{\today}

\maketitle

\emph{Introduction.}---
The quantum Hall effect is an experimentally-accessible platform to explore unconventional aspects of quantum field theory, including topological field theory \cite{wen_1995_topological,zee_2007_quantum} and emergent non-commutative geometry \cite{susskind2001quantumhallfluidnoncommutative,Polychronakos:2001mi,Hellerman:2001rj}. Furthermore, in the fractional quantum Hall regime, the Hamiltonian is dominated by interactions; a plethora of strongly-interacting phases \cite{Tsui_two,Laughlin_anomalous,Halperin_Hallcrystal,Moore_Read,jain_2007_composite} has thus been predicted to occur.

In this letter, we are interested in phases where the electronic liquid spontaneously breaks translation symmetry, forming a Wigner crystal (WC) in a strong magnetic field. While the Mermin-Wagner Theorem may ultimately destroy this order on sufficiently long scales, two-dimensional solids are visible on practical experimental scales, on which we focus.  Linear response analysis suggests that this WC has magnetophonon excitations ($\omega \sim k^2$). Such modes have been detected experimentally in the past  \cite{Andrei_observation}, albeit with unscreened Coulomb interactions which modify the dispersion to $\omega \sim k^{3/2}$.

Here, we study the dissipative dynamics of these magnetophonons using nonlinear fluctuating hydrodynamics.  In low dimensions, nonlinear fluctuations can be very strong in hydrodynamics and lead to a drastic renormalization of the dissipative dynamical scaling exponent.  The most well-known example is the breakdown of the linearized Navier-Stokes equations in $1+1$ dimensions, which flows to the Kardar-Parisi-Zhang (KPZ) universality class  \cite{Kardar, spohn_2014_nonlinear, Delacretaz:2020jis}.  However, more recently, infinitely many new such universality classes have been discovered based on ``fracton hydrodynamics".  In fracton fluids, the presence of higher multipole conservation laws gives rise to slow, subdiffusive dynamics \cite{Gromov_fracton, Feldmeier_anomalous,Morningstar_kinetically,Zhang_subdiffusion}; moreover, conserved multipole charges do not commute with spatial translations \cite{doshi_2021_vortices,Gromov:2018nbv,Huang:2023zhp}.  These effects were shown to give rise to new dynamical universality classes, distinct from KPZ, which can be sharply realized in higher-dimensional settings \cite{Glorioso:2021bif, Glorioso:2023chm}.  In the lowest Landau level (LLL), translation symmetry becomes analogous to dipole conservation \cite{Du:2021pbc}, and charge relaxes subdiffusively \cite{Hartnoll:2007ih}, so it is a natural setting to look for analogues of fracton hydrodynamics.

This letter confirms this expectation when the underlying theory transforms covariantly under Galilean boosts.  In the absence of boosts, an incoherent conductivity changes the dynamical exponent to $z=2$ which does not have an instability \cite{Baggioli:2020edn,Delacretaz:2021qqu,Delacretaz_Wignersolid}.  We show that nonlinear fluctuating hydrodynamics of the Galilean quantum Hall crystal has a similar instability to the ``dipole-and-momentum-conserving" fluid of \cite{Glorioso:2021bif, Glorioso:2023chm, Jain:2023nbf,Armas:2023ouk}.  The magnetophonon mode $\omega \sim \pm k^2 - \mathrm{i}k^z$ has dynamical critical exponent $z=4$ within linear response, which is strongly renormalized to $z\approx 3$ by relevant nonlinearities.   As many isotropic electronic systems do have approximate boost symmetry \cite{lucasdassarma}, we propose that quantum Hall crystals  could represent an experimental platform for the study of fracton-inspired dynamical universality classes.

\emph{Effective field theory.}--- We now develop a hydrodynamic effective field theory (EFT) for the quantum Hall crystal at finite temperature, setting $\hbar=e=k_{\mathrm{B}}=1$.  In the presence of a strong magnetic field $B$, we break both parity $\cP_x \cdot (x,y) =  (-x,y)$, and time-reversal $\cT \cdot t =-t $, while preserving the combination $\Theta = \cP_x \cT$.  We take the global conservation laws to be a U(1) charge $Q$, together with the two spatial components of momentum $P_{x,y}$.  In the quantum Hall regime, the momenta do not commute: \begin{align}\label{eq:PP}
    [ P_i, P_j] = \ii B  Q \epsilon_{ij},
\end{align}
 analogous to the dipole-momentum algebra \cite{Glorioso:2021bif,Glorioso:2023chm}. 
 
 We assume that energy is not a conserved quantity when building the effective field theory.  Energy relaxes diffusively, and this qualitatively changes the universality class that we will explore.  As in \cite{Glorioso:2021bif}, we will find numerically that even in systems with energy conservation, our simplified model is a good description of the dynamics on accessible time scales.  In experiments on 2d electron gases, rapid energy relaxation into (substrate) phonons may also destroy this conservation law.

 In \cite{Halperin_Hallcrystal}, it was proposed that there could be a separate conservation of U(1) charge associated to solid and fluid degrees of freedom, leading to a distinct ``Hall crystal" phase from the Wigner crystal phase.  At finite temperature, we expect thermal fluctuations lead to the conservation of only the genuine electronic charge U(1). 
 For this reason, we focus on the dynamics of the thermal Wigner crystal and simply call this the ``quantum Hall crystal". 

Since we have three continuous symmetry generators $Q$, $P_x$ and $P_y$, we follow \cite{Crossley:2015evo,Glorioso:2017fpd} and build a hydrodynamic effective field theory involving three degrees of freedom.  The role of Galilean symmetry is discussed more explicitly in \appref{app:field theory}.  Let $\sigma^A$ ($A=t,x,y$) be ``labels" for each ``cell" of the solid medium.  There is a phase variable $\varphi(\sigma^A)$ which nonlinearly realizes  the U(1) symmetry $\varphi \rightarrow \varphi + c$, along with a mapping $X^i(\sigma^A)$ ($i=1,2$) describing the physical location of each cell of the solid in space \cite{Nicolis:2013lma,Landry:2020ire}, which nonlinearly realizes translation. Since the time translation is explicitly broken, we fix $X^0 = \beta_0\sigma^t$ with $\beta_0 = 1/T_0$ the inverse temperature.  Using the coset construction  along with the non-trivial algebra \eqref{eq:PP}, we find symmetry-invariant building blocks for the Lagrangian: \begin{equation} \label{eq:singlecopyinvariants}
    E^i_A = \partial_A X^i, \;\;\; C_A = \partial_A \varphi - \frac{B}{2}\epsilon_{ij}X^j \partial_AX^i.
\end{equation}   

To build a \emph{dissipative} hydrodynamic EFT, we build a Lagrangian on the Schwinger-Keldysh contour \cite{kamenev_2011_field}, which involves doubling each degree of freedom above to $\varphi_s$ and $X^i_s$ where $s=1,2$ denotes the forward/backward time-contour.  It is convenient to take any field $\Lambda_s$ and decompose it as \begin{equation}\label{eq:def r a}
    \Lambda_r = \frac{\Lambda_1+\Lambda_2}{2}, \;\;\; \Lambda_a = \Lambda_1-\Lambda_2,
\end{equation}
where intuitively $\Lambda_r$ denotes the average value and $\Lambda_a$ denotes the stochastic fluctuations.  Naively, we simply take the symmetries of the single-copy theory and ``double" them.  However, to avoid a superfluid stiffness for the U(1) phase, we must impose an enhanced ``reparameterization" symmetry \cite{Dubovsky:2011sj,Crossley:2015evo,Glorioso:2017fpd}
\begin{align}\label{eq:varphi repara}
        \varphi_r(\sigma^A)\to \varphi_r(\sigma^A) + \alpha(\sigma^I),
\end{align}
which can be understood as a consequence of unbroken weak symmetry in the thermal state \cite{Akyuz:2023lsm,Huang:2024rml}; here $I=x,y$ denotes only the spatial components of $\sigma^A$.  Note that $\partial_t \varphi_r$ is the chemical potential of the U(1), which can be nonzero, which is why $\alpha(\sigma^I)$ only depends on the spatial coordinates.   In contrast, the quantum Hall crystal does spontaneously break (weak) translation symmetry, and so we do not have such a reparameterization symmetry for $X^i$; the homogeneity of the crystal does allow us to freely set the origin of the $\sigma^A$ coordinates: \cite{Nicolis:2013lma,Landry:2020ire}
\begin{align}\label{eq:sigma repara}
    \sigma^I\to \sigma^I +c^I.
\end{align}
A crystal also spontaneously breaks the other symmetries, like the rotational and Galiean boost symmetries, but that will not generate additional Goldstones \cite{Watanabe_redundancies,Nicolis:2015sra,Landry:2019iel}.  In what follows, we will drop the ``$r$" subscript on fields, as the eventual EFT is expanded in powers of ``$a$" fields. 

It is useful to work in the physical spacetime. 
From \eqref{eq:singlecopyinvariants} we can define $E^I_i$ as the inverse of $E^i_I$ (e.g. $E^J_i E^i_I = \delta^J_I$) and use this ``vielbein" to convert to $E^I_\mu$ and $C_\mu$. The time-derivatives of these fields are identified with physical observables: velocity $u^\mu$ and chemical potential $\mu$ are defined by \begin{equation}\label{eq:r inv}
    E^\mu_t \equiv \beta_0 u^\mu, \;\;\; C_t \equiv \beta_0\mu.
\end{equation}
Notice that this process implements the solid Josephson relation: 
\begin{equation*}
    0 = \frac{\ud X^i}{\ud X^0} = \frac{\partial \sigma^A}{\partial X^0} E^i_A = u^i + \partial_0 \sigma^I E^i_I,
\end{equation*} 
which implies that \begin{equation} \label{eq:josephson relation}
    \partial_0 \sigma^I = -E^I_i u^i \approx -  \delta^I_i u^i,
\end{equation}
which locks the local velocity to the motion of the solid, and we used $\langle\sigma^I(x)\rangle = x^i \delta^I_i$ and ignored higher-order nonlinear terms.

The final thing we must implement is $\Theta$-symmetry, which leads to a generalized Kubo-Martin-Schwinger (KMS) symmetry.  In the classical limit of interest to us for hydrodynamics, \cite{Glorioso:2017fpd} 
\begin{align}\label{eq:kms}
    \tilde\Lambda(\Theta x) = \Theta \Lambda(x),\;\; \tilde \Lambda_a(\Theta x)  = \Theta\left(\Lambda_a + \ii \cL_{\beta} \Lambda\right)(x),
\end{align}
where $\cL_{\beta}$ is the Lie derivative along $\beta^\mu = \beta_0 u^\mu$, and $\Lambda$, $\Lambda_a$ are the invariant building blocks: for $a$-fields these are (at lowest non-trivial order in $a$-fields)
\begin{align}\label{eq:a inv}
    E^i_{a,\mu} &= \p_\mu X^i_a, \;\; C_{a,\mu} = \p_\mu \varphi'_a - B   \delta_\mu^i X^j_a \epsilon_{ij},
\end{align}
where we defined $\varphi_a' \equiv \varphi_a - \frac{1}{2}B X_a^i X^j \epsilon_{ij}$.
For a system to approach thermal equilibrium, the effective action should have KMS symmetry.

\emph{Hydrodynamics.}---  At the ideal level, the most general effective Lagrangian is given by
\begin{align}\label{eq:ideal L}
    \cL_{\mathrm{ideal}} =~& \left(p \delta^\mu_i  + \pi_i u^\mu  +\kappa^\mu_j \p_i\sigma^j \right) E^i_{a,\mu}+n u^\mu C_{a,\mu},
\end{align}
where the prefactors are functions of the invariant $r$-fields, and we define $\sigma^i = \delta^i_I\sigma^I$. 
In order for it to be invariant under the KMS transformation \eqref{eq:kms}, we demand 
\begin{align}\label{eq:eos}
    \delta p = \pi_i \delta u^i - \kappa^\mu_i \delta(\p_\mu \sigma^i) + n\delta \mu ,
\end{align}
where we used $\beta_0 u^\mu \p_\mu \sigma^i(x) = \delta^i_I\ud\sigma^I/\ud \sigma^t =0$, 
so that the change $-\ii T_0(\tilde \cL_{\mathrm{ideal}} - \cL_{\mathrm{ideal}}) = \p_{\mu}(p u^\mu)$ becomes a total derivative. This gives rise to an entropy current $s^\mu = \beta_0 p u^\mu$ that satisfies $\p_\mu s^\mu=0$ \cite{Glorioso:2016gsa,Glorioso:2017fpd}. \eqref{eq:eos} gives the equation of state with $p$ the thermodynamic pressure, $\pi_i$ the momentum density, $\kappa^\mu_i$ the strain density, and $n$ the charge density. 
Expanding the pressure to the quadratic order in $\delta n = n - n_0$, $\phi^i = \sigma^i - x^i$ and $u^i$, we find 
\begin{align}\label{eq:pressure}
    p \approx~& p_0 + n_0\delta \mu +  \frac{\delta n^2}{2\chi}+\frac{\pi_i^2}{2 m n_0} \nonumber\\
    &- \frac{1}{2} a^{ijkl} \p_i \phi_j \p_k \phi_l + \frac{1}{2} b (\p_0 \phi_i)^2,
\end{align}
where $\chi = \p n /\p\mu$ and $mn_0 = \pi_i/u_i$ are the charge and momentum susceptibility, and $a^{ijkl}$ is the modulus tensor given by\footnote{Since the rotational symmetry is broken spontaneously, $a^{ijkl} = a^{jikl} = a^{ijlk}$.}
\begin{align}\label{eq:atensor}
    a^{ijkl} &= a_1 \delta^{ij}\delta^{kl}+a_2 \delta^{i<k} \delta^{l>j},
\end{align}
with $A^{<ij>}=A^{ij}+A^{ji}-\delta^{ij}A^{kk}$; $a_{1,2}$ are the bulk and shear modulus, respectively.
Thermodynamic stability requires that $\chi,m,b,a_{1,2}\geq 0$. Upon variation of $a$-fields, we find the following equations of motion at leading order in $\omega$ and $k$, as
\begin{subequations}\label{eq:WI linear}
\begin{align}
    \p_i p +\p_k(\kappa^k_j\p_i\sigma^j)- B \epsilon_{ij} n_0 u^j &=0,\label{eq:momentum eom}\\
    \p_0 \delta n + n_0 \p_i u^i &=0.
\end{align}
\end{subequations}
$\p_0\pi^i$ is irrelevant compared to the Lorentz force term.

If we turn on an external electric field: $\mu\to \mu - E_i x_i$, \eqref{eq:momentum eom} implies  the Hall current
\begin{align}\label{eq:Hall current}
    J^i \equiv n_0 u^i =  \nu \epsilon^{ij}E_j,
\end{align}
where $\nu = n_0/B$ is the filling fraction.  Using \eqref{eq:josephson relation} and \eqref{eq:pressure}, we can rewrite the Ward identities \eqref{eq:WI linear} as
\begin{subequations}\label{eq:WI ideal 2}
\begin{align}
    B n_0 \epsilon_{ij}\p_0\phi^j   + \chi^{-1} n_0 \p_i\delta n + a^{ijkl}\p_j\p_k \phi_l  &=0 ,\label{eq:phonon eom}\\
    \p_0[\delta n - n_0 \p_i \phi^i] &=0. \label{eq:deltanfixed}
\end{align}
\end{subequations}
Thus, charge fluctuations are identified as the divergence of the lattice deformation at the ideal level.
Solving it, we find a quadratic-dispersing magnetophonon $\omega=\pm ck^2$ with  $c = (B n_0)^{-1}\sqrt{a_2(a_1+a_2+\chi^{-1}n_0^2)}$. 


We turn to dissipaive corrections. The KMS-invariant term $\cL^{(1)}\sim \ii T_0 \sigma_0 C_{a,i}\left(C_{a,i} + \ii \beta_0 (\p_i \mu - B \epsilon_{ij}u^j)\right)$  gives a finite relaxation rate $\sigma_0 B^2/m$ for the momentum density on top of the cyclotron dynamics suppressed from \eqref{eq:momentum eom}, as $\sigma_0$ plays the role of an incoherent conductivity \cite{Hartnoll:2007ih}. Such a term will give the magnetophonon a quadratic decay rate: $\omega\sim \pm k^2 - \ii k^2$, along with a finite longitudinal conductivity \cite{Baggioli:2020edn,Delacretaz:2021qqu,Delacretaz_Wignersolid}. 

 Two-dimensional electron gases with an isotropic Fermi surface exhibit an approximate Galilean symmetry at low temperature \cite{lucasdassarma}.  
 Following \cite{Glorioso:2023chm}, Galilean symmetry implies that only $E_{a,0i}+m^{-1}C_{a,i}$ is an invariant building block (both $E_{a,0i}$ and $C_{a,i}$ shift by a constant under boost): see \appref{app:field theory}.  Dissipative terms coupled to $E_{a,0i}$ can be absorbed into fluid frame transformations.  Therefore, for the Galilean-invariant system, the important dissipative terms are proportional to  $U_{a,ij}\equiv \p_i C_{a,j} + B\epsilon_{jk} E^k_{a,i} = \p_i\p_j \varphi'_a$ together with $E_{a,ij}$.
The most general dissipative KMS-invariant Lagrangian is found as
\begin{align}\label{eq:diss L}
    \cL_{\mathrm{diss}} =~& \ii T_0 s^{ijkl}U_{a,ij}\left(U_{a,kl} + \ii \beta_0 \p_k \p_l \mu\right)\nonumber\\
    &+\ii T_0 t^{ijkl}E_{a,ij}\left(E_{a,kl} + \ii \beta_0 \p_k u_l\right)
\end{align}
where the invariant tensors $s^{ijkl},t^{ijkl}$  have the same decomposition as \eqref{eq:atensor} with $s_{1,2},t_{1,2}\geq 0$ due to unitarity \cite{Crossley:2015evo}. There is also a $\Theta$-even Lagrangian that is non-dissipative but higher-order compared to the ideal Lagrangian. Keeping the leading order scaling as wave number $k\rightarrow 0$, the normal modes are a magnetophonon ($\omega = \pm c k^2 - \ii \Gamma k^4$) and a subdiffusive charge mode ($\omega =-\ii D  k^4$).

\emph{Instabilities of hydrodynamics.}---So far, we considered the linear response theory. The relevant nonlinearity arises from expanding the elastic modulus and charge susceptibility in terms of small deviations $\p_i \phi^i$ (or, equivalently, $\delta n$): $a^{ijkl}\approx a^{ijkl}_0+\lambda^{ijkl} \p_{i'} \phi^{i'},\; \chi^{-1} \approx \chi^{-1}_0 +\lambda' \p_{i'} \phi^{i'}$. 
Converting from the KMS-invariant EFT to a nonlinear Langevin equation, we find
\begin{subequations}
    \label{eq:eom nonlinear}
\begin{align}
    B &n_0 \epsilon_{ij}\p_0\phi^j  + \chi^{-1}_0n_0\p_i \delta n + a^{ijkl}_0\p_j\p_k \phi_l+t^{ijkl}\p_j\p_k\p_0 \phi_l\nonumber\\
    &+\lambda'n_0\p_i (\p_j\phi^j\delta n)+ \lambda^{jikl}\p_j (\p_{i'} \phi^{i'}\p_k \phi_l)=\p_j \xi^{ij},  \\
    &\p_0 \delta n - n_0 \p_0 \p_i \phi^i + \chi^{-1}_0(s_1+s_2)\p^4 \delta n  = \p_i\p_j \tau^{ij},&
\end{align}\end{subequations}
where the noise fields satisfy $\langle\xi^{ij}(x)\xi^{kl}(0) \rangle = 2T_0 t^{ijkl} \delta^{(d)}(\vect x)\delta(t)$, $\langle\tau^{ij}(x)\tau^{kl}(0) \rangle = 2T_0 s^{ijkl} \delta^{(d)}(\vect x)\delta(t)$. At the dissipative fixed point $\omega\sim k^{4}$, the noise field scales as $\xi^{ij},\tau^{ij}\sim k^{(d+4)/2}$. To match the
scaling to the dynamical terms with time derivatives, we find $\delta n \sim k^{d/2},\;\phi^i\sim k^{d/2-1}$. By comparing the nonlinear terms with the noise field, we have
\begin{align}\label{eq:lambda scaling}
    \lambda^{ijkl}\sim \lambda'\sim k^{(4-d)/2},
\end{align}
meaning the nonlinearities are relevant when $d<4$. As a consequence, we expect the
true IR fixed point to have anomalous dissipative scaling $\omega \sim k^z$ with $z\leq 4$. As we do not expect the thermodynamic fields to exhibit renormalized scaling, which can be understood as a consequence of the IR fixed point having the same statistical steady state (by KMS symmetry \cite{Huang:2023eyz}), we can crudely estimate the fixed point to be $z^* \approx d/2+2$ by treating $\lambda$, rather than $s/t$, as marginal.   In $d=2$, we find $z^* \approx 3$.  Nonlinearities cubic in $\p_i \phi^i$ in e.g. the pressure are marginal at tree-level, in contrast to \eqref{eq:lambda scaling}.


Note that based on \eqref{eq:pressure}, the $k$-integral of the phonon susceptibility 
\begin{align}\label{eq: chi phonon}
    \chi_{\phi^i\phi^i}\propto 1/k^2
\end{align}
has an IR divergence \cite{Mermin_crystalline}. This should not be regarded as a contradiction to the non-equilibrium universality class: it is a property of the equilibrium state. Moreover, a slightly long-range interaction is enough to suppress the IR divergence while retaining the instability of linearized hydrodynamics to nonlinear fluctuations (see \appref{app:field theory}).

\begin{figure}[t]
\includegraphics[width=1.\linewidth]{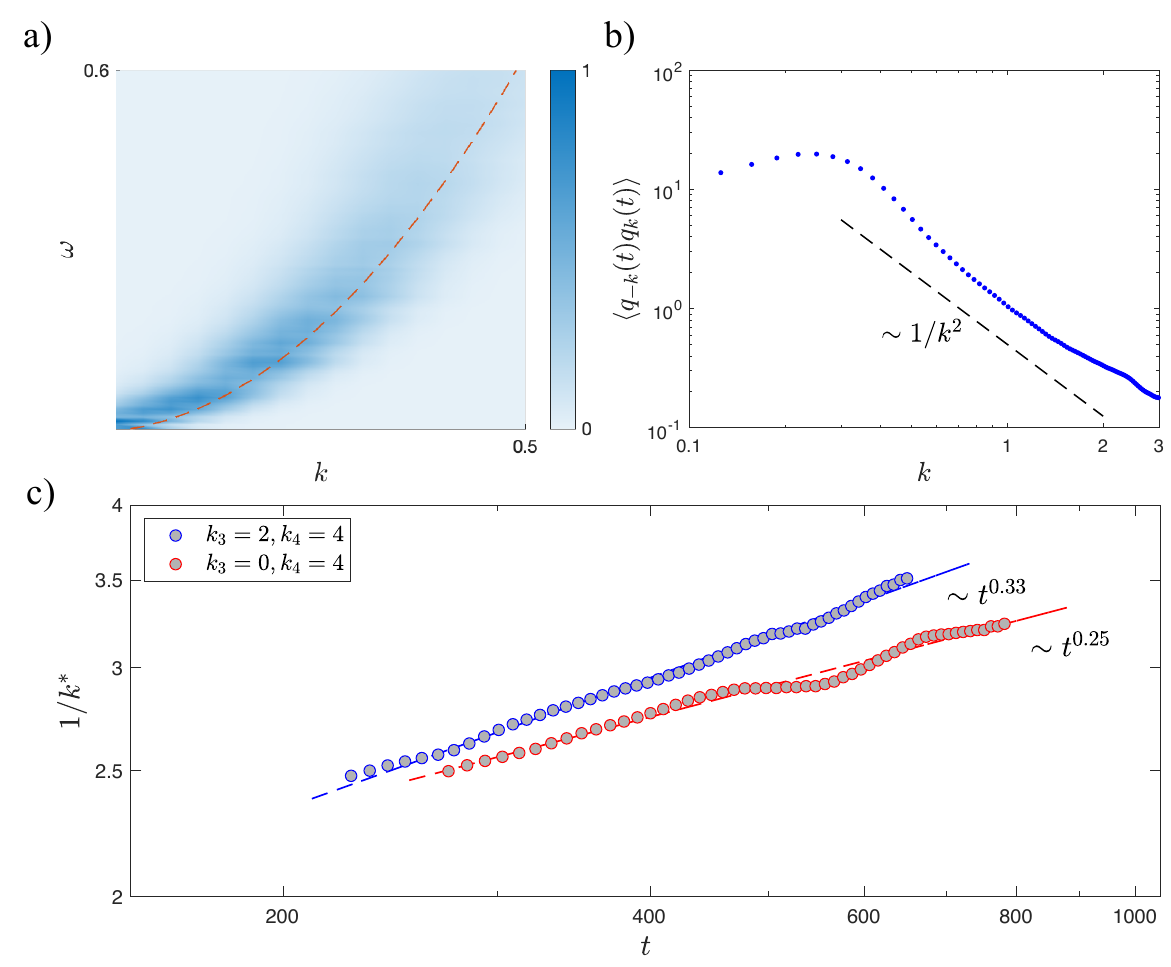}
\caption{\justifying Numerical simulation of the classical dynamics generated by Hamiltonian \eqref{eq:HLLL} and Poisson bracket \eqref{eq:pb}. a) Normalized value of the temporal Fourier transform of $\langle q_{-k}(0)q_{k}(t)\rangle$, showing the quadratic-dispersing magnetophonon. b) Equal-time correlation function $\langle q_{-k}(t)q_{k}(t)\rangle$ at late time $t\to \infty$, consistent with theoretical $1/k^2$ prediction. c) Temporal dependence of $1/k^*$, showing anomalous scaling $1/z\approx 0.33$ for a relevant nonlinear potential ($k_3=2,k_4=4$). $k^*$ is the wavevector at which the curve in b) develops the maximum. For a marginal nonlinear potential ($k_3=0,k_4=4$), the scaling is consistent with the linear response theory ($1/z = 1/4$). The simulations were done on a $200\times 200$ square lattice with periodic boundary conditions, averaged over 32 realizations of random initial conditions. The data in c) has been smoothed by averaging over adjacent time steps; we have verified that this does not affect the scaling to the second decimal place.
}
\label{fig:numerics}
\end{figure}

\emph{Numerical simulations.}---We have tested the predictions of our EFT in a microscopic model consisting of particles on a square lattice, which forms a crystal. In the lowest Landau level limit, the phase space is given by a ``non-commutative geometry":  letting $q_i(n)$ denote the displacement of the particle at position $n=(n_x,n_y)$ in the lattice in the $i^{\mathrm{th}}$ direction, we have Poisson bracket
\begin{align}\label{eq:pb}
    \{q_i( n),q_j( n')\} = \epsilon_{ij}\delta_{ n, n'},
\end{align}
The magnetic field strength has been set to unity in these units. We write an effective low-energy Hamiltonian
\begin{align}\label{eq:HLLL}
    H_{\mathrm{LLL}} = \sum_{n_x,n_y} V(\Delta_i q^i(\vect n)) + \frac{1}{2}(\epsilon^{ij}\Delta_i q_j(\vect n))^2,
\end{align}
where $\Delta_i q^i(\vect n)\equiv q^x(n_x,n_y) - q^x(n_x - 1, n_y) + q^y(n_x,n_y) - q^y(n_x, n_y-1)$ is the discrete divergence. We take an anharmonic potential $V(x) = \frac{1}{2}x^2 + \frac{1}{3}k_3 x^3+\frac{1}{4} k_4 x^4 +\ldots$. The quadratic potentials in \eqref{eq:HLLL} encode the bulk and shear modulus of the crystal, while the cubic potential $k_3$ captures to the relevant nonlinearity $\lambda^\prime$ from \eqref{eq:eom nonlinear}, 
and the quartic potential $k_4$ corresponds to the next leading order, which are marginal at tree level.
Using the Poisson bracket \eqref{eq:pb}, we can derive the Hamilton's equations for each displacement.  Although energy is formally conserved in this model, we will see that the numerics are compatible with our energy-free EFT on accessible time scales, as in the dipole-momentum theory \cite{Glorioso:2021bif}.  In Appendix \ref{app:galileanH}, we show that this model is Galilean-covariant.  Therefore, we expect to observe the new fixed point with $z\approx 3$ in numerical simulations with $k_3 \ne 0$.

We simulated this model using $4\times 10^4$ particles. The divergent phonon susceptibility \eqref{eq: chi phonon} at $k\to 0$ is observed in \figref{fig:numerics}b through the equal-time correlation function $\langle q_{-k}(t) q_{k}(t) \rangle$, where $\langle...\rangle$ denotes average over initial conditions, and $q_{k}$ is the discrete Fourier transform of $q(n)$. 
In \figref{fig:numerics}a, we show a quadratic dispersing magnetophonon as the peaks of the temporal Fourier transform of unequal-time
correlation functions $\langle q_{-k}(0) q_{k}(t) \rangle$. The peaks get broadened due to the dissipation caused by nonlinear potentials. 
To characterize the time scale associated to dissipation, we study the temporal dependence of the wavevector $k^*$ at which $\langle q_{-k}(t) q_{k}(t) \rangle$ develops a maximum, after starting from random initial conditions. A dissipative mode $-\im \omega \sim  k^z$ causes modes with $k>k^*$ to decay in time $t\sim k^{*-z}$, implying that we should see 
\begin{align}
    1/k^* \sim t^{1/z}.
\end{align}
In \figref{fig:numerics}c), we extract the scaling to be $z\approx 4$ when $k_3=0$,  and $z\approx 3$ when $k_3\ne 0$. This is consistent with our prediction that only $k_3$ drives a flow to a new dynamical universality class, which we find has $z\approx 3$.

\emph{Experimental probes.}--- Unlike the dipole-momentum fixed point, we now argue that the unconventional magnetophonon decay predicted above may be accessible in experiments.  The magnetophonon can couple to the electromagnetic probe through electron-electron interactions, as the Wigner crystal of electrons is charged. We propose using microwave impedance microscopy (MIM) to detect the magnetophonon and its dissipation. MIM utilizes a reflected microwave signal to measure the density-density correlation function $\chi_{nn}(k,\omega)$ \cite{barber_2021_microwave,Taige_mim}. Following \cite{Taige_mim}, the tip-sample admittance, proportional to the reflected signal, is given by \begin{equation}
    Y(\omega)\approx - \ii \omega L^{-1} \sum_n G_r(k_n)^2 \chi_{nn}(k_n,\omega),
\end{equation}
where $L$ is the sample perimeter, $G_r(k)$ is the Fourier transform of the Coulomb interaction with $r$ the linear distance between the tip and the sample and the  discrete summation is over normal modes in the device geometry. Using \eqref{eq:deltanfixed},  we estimate
\begin{align}
    \re Y(\omega)&\sim  \omega \notag \sum_n G_r(k_n)^2  k_n^2 \im \chi_{\phi_i \phi_i}(k_n,\omega)\\
    &\sim \sum_n G_r(k_n)^2 \frac{\omega^2 k_n^{4+z}}{(\omega^2 - c^2 k_n^4)^2+\Gamma^{'2}\omega^2 k_n^{2z}},
\end{align}
Due to the discrete wavevector, $\re Y(\omega)$ will develop a series of peaks at $\omega_{\mathrm{MIM}} = c k_n^2$ with the broadening of the peak proportional to $\Delta \omega \sim \Gamma'  k_n^{z}$. Hence, by measuring the width of the peak, we can extract the dynamical exponent $z$. 

Suppose the crystal has a charge density $n_0\approx 10^{9}/\mathrm{cm}^2$, a typical elasticity modulus $a_0 \approx 10^{-9}$N/m, and a phonon velocity in the absence of external fields as $c_0 \sim \sqrt{a_0/m_e n_0} \approx 10^4\mathrm{m}/\mathrm{s}$ \cite{Zhao_dynamic}. Given a magnetic field $B\approx 0.4 \mathrm{T}$, the magnetophonon ``velocity'' is around $c \sim c_0^2/\omega_c \approx 10^{-2}\mathrm{m}^2/\mathrm{s}$, where $\omega_c$ is the cyclotron frequency. For $L = 10\mu$m, the first peak of $\re Y$ will occur near $\omega_{\mathrm{MIM}} \approx 1 \mathrm{GHz}$ which is within the range of microwave. We then estimate the dissipation at $z=3$ using thermodynamic variables. By dimensional analysis, we find
\begin{align}\label{eq:ratio}
    \frac{\Delta\omega}{\omega_{\mathrm{MIM}}} \sim \frac{\Gamma'}{c}k_n \sim \sqrt{\frac{T_0}{a_0}} \tilde\lambda k_n,
\end{align}
where $\tilde \lambda$ is the dimensionless nonlinear coupling. Let us take $T_0 = T_\mathrm{m}$ to be the melting temperature, then, with a filling fraction $\nu = 0.1$, $T_\mathrm{m}\approx 1\; \mathrm{K}$ \cite{chen_2006_melting}. Together with $\tilde\lambda\sim O(1)$, this yields $\Delta\omega/\omega_{\mathrm{MIM}} \approx 0.1$, which is consistent with a sharp and visible peak.

\emph{Outlook.}---We have argued that magnetophonons in a quantum Hall crystal are unstable against nonlinear perturbations.  This represents a promising route to discover new universality classes, inspired by fracton hydrodynamics, in experiment.  This phenomenon should occur, in principle, in high-quality GaAs heterostructures as studied in \cite{Andrei_observation}.  



The Tkachenko mode in a rotating vortex lattice \cite{sonin_2014_tkachenko} also exhibits a similar dispersion relation to the magnetophonon.  It would be interesting to investigate whether this theory also has a breakdown of linearized dissipative hydrodynamics, using the EFT for such systems \cite{Du:2022xys}. 



\emph{Acknowledgements.}---We acknowledge useful discussions with Liang Fu, Blaise Gout\'eraux, Leo Radzihovsky, Dam Thanh Son, Taige Wang, and Isabella Zane.
This work was supported by the Gordon and Betty Moore Foundation's EPiQS Initiative under Grant GBMF10279 and the National Science Foundation under CAREER Grant DMR-2145544.

\bibliography{QHC}

\onecolumngrid

\newpage

\begin{appendix}

\section{Hydrodynamic effective field theory}\label{app:field theory}

In this appendix, we present a detailed construction of the hydrodynamic effective field theory of the Galilean-invariant quantum Hall crystal, following \cite{Crossley:2015evo,Glorioso:2017fpd}. 

\subsection{Coset construction}
In the presence of a magnetic field, we have a modified Galilean algebra
\begin{subequations}\label{eq:algebra app}
\begin{align}
    [P_i,P_j]&=\ii B Q \epsilon_{ij},\\ [K_i,P_j] &= \ii m Q \delta_{ij},\\
    [K_i,P_0] &= -\ii(P_i - (B/m) \epsilon_{ij}K_j), \label{eq:K P0 app}
\end{align}
\end{subequations}
where $K_i$ is the boost generator and $m$ is the single particle mass.  The algebra is closed, and commutators not listed are zero. 

To motivate this algebra, let us consider the classical Hamiltonian describing a quadratically-dispersing particle (with Galilean boost symmetry) placed in am magnetic field in two spatial dimensions:
\begin{equation}
    H = \frac{1}{2m}\left(p_i - \frac{B}{2} \epsilon_{ij}x_j\right)^2
\end{equation}
 with canonical Poisson brackets $\{x_i,p_j\} = \delta_{ij}$ along with $\lbrace x_i,x_j\rbrace = \lbrace p_i,p_j\rbrace = 0$.  \eqref{eq:algebra app} follows from identifying $P_0 = -H$, $K_i=m x_i$, and \begin{equation}
     P_i = p_i + \frac{B}{2}\epsilon_{ij}x_j. \label{eq:Pivspi}
 \end{equation}
 The motivation for \eqref{eq:Pivspi} is that in the limit $m\rightarrow 0$, the degrees of freedom $p_i$ and $x_i$ are not independent; ordinarily this is seen by considering Dirac brackets and reducing the effective phase space from $(x,y,p_x,p_y)$ to $(x,y)$, with the latter called the non-commutative plane; indeed, we considered this limit when studying Hamiltonian dynamics in the main text.  Moreover, we see that $[P_i,P_0]=0$ with the choice \eqref{eq:Pivspi}, and this makes sure that the algebra can reduce to \eqref{eq:PP} by explicitly breaking the Galilean boost.   For us, we wish to make an analogy with the ordinary coset construction for a solid in which $P_{x,y}$ generate translations, and since \eqref{eq:Pivspi} does not vanish in the limit $m\rightarrow 0$ (due to the sign discrepancy with the argument of $H$), it is a reasonable choice.  Notice that the $m\rightarrow 0$ limit of \eqref{eq:algebra app} is naively sick due to the $m^{-1}$ in \eqref{eq:K P0 app}, and we will return to this point shortly.

The dynamical degrees of freedom are those that nonlinearly realize the translational, U(1) and Galilean boost symmetries. For momentum conservation, we introduce a set of dynamical coordinates $X^i(\sigma)$ that nonlinearly realize the translational symmetries $ P_i$: $X^i(\sigma) \to X^i(\sigma)+\xi^i$, where $\xi^i$ is constant. For charge conservations, we introduce $\varphi(\sigma)$ and that transforms as $Q: \varphi(\sigma)\to \varphi(\sigma)+a$. For Galilean boost, we introduce the Goldstone $\eta^i(\sigma)$ that nonlinearly realize it: $K_i:\eta^i(\sigma)\to \eta^i(\sigma) +b^i$. In a normal fluid or solid, we do not have the boost Goldstone as the degree of freedom, because it is redundant to the fluid velocity \cite{Glorioso:2023chm}.  We will shortly show that the same effect arises  here. 
The coset construction can be used to systematically and efficiently find the invariant building blocks under global (spontaneously broken) symmetries  \cite{Nicolis:2013lma,Nicolis:2015sra,Landry:2019iel}. 
To describe non-equilibrium dynamics, we put the fields on the Schwinger-Keldysh contour ($s=1,2$). The coset elements are parametrized by ($s=1,2$) 
\begin{align}
    g_s(\sigma) =e^{\ii \beta_0\sigma^t  P_0} e^{\ii X^i_s(\sigma) P_i} e^{\ii \eta^i_s(\sigma) K_i} e^{\ii \varphi_s(\sigma) Q}.
\end{align}
The Maurer-Cartan 1-form is given by $g^{-1}_s\p_A g_s =  \ii E^\mu_{s,A}  P_\mu + \ii C_{s,A} Q +\ii V^i_{s,A} K_i  $ with
\begin{subequations}\label{eq:blocks}
\begin{align}
        E^i_{s,A} & = \p_A X^i_s-\beta_0 \delta_{A,t}  \eta^i_s,\\ C_{s,A}  &= \p_A \varphi_s - \frac{B}{2} \p_A X^i_s X^j_s \epsilon_{ij}+m \p_A X^i_s \eta_{s,i} - \frac{m\beta_0}{2} \delta_{A,t}(\eta^i_s )^2,\\
        V^i_{s,A}& =\p_A \eta^i_s - \delta_{A,t}\beta_0\frac{B}{m} \epsilon^{ij} \eta_{s,j},
\end{align} 
\end{subequations}
and $E^0_{s,A} = \beta_0 \delta_{A,t}$. The invariant $r$-fields are
\begin{subequations}\label{eq:r inv app}
    \begin{align}
        E^i_{\mu} &= \delta^i_\mu - \delta_{\mu,0}  \eta^i,\\
        C_t & = \p_t \varphi - \frac{B}{2}\beta_0 u^i X^j\epsilon_{ij} + m\beta_0 u^i \eta_i - \frac{m\beta_0}{2}  \left(  \eta^i\right)^2,\\
        V^i_\mu & = \p_\mu \eta^i - \delta_{\mu,0}\frac{B}{m}\epsilon^{ij}\eta_{j},
    \end{align}
\end{subequations}
with $\p_t X^\mu = \beta_0 u^\mu$.
The $a$-fields become (at lowest non-trivial order)
\begin{subequations}\label{eq:a inv app}
    \begin{align}
        E^i_{a,\mu} &= \p_\mu X^i_a- \delta_{\mu,0}  \eta^i_a, \\ C_{a,\mu} &= \p_\mu \varphi'_a -B \delta^i_\mu\epsilon_{ij}X^j_a+ m \delta^i_\mu \eta_{a,i} +  m\eta_i E^i_{a,\mu},\\
        V^i_{a,\mu}&= \p_\mu \eta^i_a - \delta_{\mu,0}\frac{B}{m}\epsilon^{ij}\eta_{a,j}
    \end{align}
\end{subequations}
with $\varphi'_a=  \varphi_a - \frac{1}{2}B X^i_a X^j\epsilon_{ij}$.

The fact that not all of the dynamical degrees of freedom are Goldstone modes lead to additional constraints---these are the reparameterization symmetries \eqref{eq:sigma repara} and \eqref{eq:varphi repara}. 

The unitarity constraint of the underlying microscopic motion reflects in an effective action through \cite{Crossley:2015evo}
\begin{align}
    S[\Lambda,\Lambda]=0,\quad S[\Lambda_1,\Lambda_2] = - S^*[\Lambda_2,\Lambda_1],\quad \im S[\Lambda_1,\Lambda_2] \geq 0,
\end{align}
where the action $S$ depends on two copies of the fields.
Note that the action can be complex-valued, which manifests the dissipative nature of the non-equilibrium dynamics. The microscopic dynamics in quantum Hall crystal is not invariant under the time-reversal symmetry, but $\Theta = \cP_x \cT$. Under $\Theta$, the variables change as
\begin{align}
    x^\mu \to (-x^0,-x^1,x^2),\quad u^\mu \to (u^0,u^1,-u^2),\quad
    \eta^i \to (\eta^1,-\eta^2),\quad
    \varphi \to -\varphi.
\end{align}
For the system to approach an effective thermal state set by the inverse temperature $\beta_0$, there is a $\mathbb{Z}_2$ symmetry called the dynamical KMS symmetry obeyed by the effective action \cite{Crossley:2015evo,Glorioso:2017fpd}
\begin{align}
    S[\Lambda_1,\Lambda_2] = S[\tilde \Lambda_1,\tilde \Lambda_2],\quad \tilde \Lambda_1 (\Theta \sigma) = \Theta \Lambda_1(\sigma),\quad  \tilde \Lambda_2 (\Theta \sigma) = \Theta \Lambda_2(\sigma^t+\ii \beta_0,\sigma^I)
\end{align}
To take the classical limit, we introduce the decomposition $\Lambda_1 = \Lambda + \frac{1}{2}\hbar\Lambda_a$ and $\Lambda_2 = \Lambda - \frac{1}{2}\hbar\Lambda_a$, and take $\hbar\to 0$. Equivalently, we will omit the factor $\hbar$ and work to the leading order in $a$-fields. It is convenient to work in the physical spacetime in the classical limit to connect with the standard form of hydrodynamic theory. The pull-back from the  physical spacetime to the fluid spacetime is carried by $e_{s,A}^\mu = \p_A X_s^\mu$. Now, if $\Lambda_{s,A}(\sigma) = e_{s,A}^\mu \Lambda'_{s,\mu}(X_s)$, we find $\Lambda_{a,A} = e_A^\mu \Lambda_{a,\mu}\equiv    e_A^\mu(\Lambda'_{a,\mu} + \cL_{X_a} \Lambda'_{\mu})$. Since $\tilde X_a^\mu(\Theta \sigma) =\Theta( X_a^\mu + \ii \beta_0 \p_t X_a^\mu  )(\sigma) $ in the classical limit, we find the dynamical KMS symmetry in the physical spacetime and classical limit becomes \eqref{eq:kms}.

An effective Lagrangian to the leading order in $a$-fields is given by
\begin{align}\label{eq: eff L}
    \cL = \hat{T}^\mu_i E^i_{a,\mu}  + J^\mu C_{a,\mu} + W^\mu_i V^i_{a,\mu}.
\end{align} 
By variation of the $a$-fields, we get the Ward identities, to the leading order in fields,
\begin{subequations}
    \begin{align}
        \p_\mu T^\mu_i+B J^j\epsilon_{ji} &=0,\\
        \p_\mu J^\mu & = 0,\\
        \p_\mu W^\mu_i+\frac{B}{m}W^0_j\epsilon_{ji}+T^0_i-mJ_i &=0,\label{eq:boost WI}
    \end{align}
\end{subequations}
 which correspond to the momentum, charge and boost conservation equations respectively. In the above equation, we defined the momentum density as
 \begin{align}
     T^\mu_i\equiv \hat{T}^\mu_i +  mJ^\mu \eta_i.
 \end{align}
 
Recall that the last commutator \eqref{eq:K P0 app} contains the Galilean boost generator on the right hand side.  The dynamical consequence of this can now be clearly seen in \eqref{eq:boost WI}: the ``boost density" $W^0_i$ undergoes oscillations at the cyclotron frequency $\omega_{\mathrm{c}} = B/m$.  This is capturing Kohn's theorem \cite{Kohn_cyclotron}, and implies that the boost will not be a slow hydrodynamic degree of freedom.

 Moreover, the boost conservation equation \eqref{eq:boost WI} in the hydrodynamic limit becomes a constraint that the momentum density
is equal to the mass current:
 \begin{align}\label{eq:momentum current identity}
     T^0_i = m J_i.
 \end{align}
 This implies that the boost Goldstone $\eta^i$ is not a dynamical degree of freedom, and we should integrate it out. Further, the first order dissipative Lagrangian contains the following term 
 \begin{align}
     \cL^{(1)}\sim \ii T\sigma_0 C_{a,i}(C_{a,i} + \ii \cL_{\beta}C_i)\supset - \sigma_0 m^2 \eta_{a,i}\p_0 \eta_i,
 \end{align}
where $\sigma_0$ is the incoherent conductivity. The leading order equation of motion for the boost Goldstone becomes
\begin{align}
    \hat T^0_i -m J_i +c_0\eta_i+\sigma m^2 \p_0\eta_i = O(\partial^2),
\end{align}
with $c_0$ some thermodynamic coefficients.
Thus, the boost Goldstone will be relaxed at a timescale $\propto \sigma_0$ (multiplied by other ``thermodynamic" coefficients) to 
\begin{align}\label{eq:eta constraint}
    \eta_i\approx 0.
\end{align}
Under \eqref{eq:eta constraint}, the invariant $r$-fields \eqref{eq:r inv app} reduces to \eqref{eq:r inv} in the main text.

Integrating out $\eta_i$ in the effective action amounts to removing $\eta_{a,i}$, which can be achieved by introducing the $\eta_{a,i}$-independent variables
\begin{subequations}
    \begin{align}
        E^{ \prime}_{a,0i} & \equiv E_{a,0i} + m^{-1} C_{a,i} = \p_0 X^i_a +m^{-1}\p_i \varphi'_a - (B/m)\epsilon_{ij} X^j_a,\\
        U_{a,ij}&\equiv \p_i C_{a,j} - m V_{a,ij}+B\epsilon_{jk} E^k_{a,i} = \p_i\p_j \varphi'_a,
    \end{align}
\end{subequations}
and neglecting $V^i_{a,\mu}$.
The KMS transformations of these $a$-fields, together with $E_{a,ij}$, are given by, in the classical limit and physical spacetime,
\begin{subequations}
    \begin{align}
        \tilde E_{a,ij}(\Theta x) & = \Theta\left(E_{a,ij} + \ii \beta_0 \p_i u_j \right),\\
        \tilde E'_{a,0i}(\Theta x) & = \Theta \left( E'_{a,0i} + \ii \beta_0 (\p_0 u_i - (B/m)\epsilon_{ij} u^j + m^{-1}\p_i \mu)\right),\\
        \tilde U_{a,ij}(\Theta x) &= \Theta\left(U_{a,ij} + \ii \beta_0 \p_i \p_j \mu\right).
    \end{align}
\end{subequations}


\subsection{Ideal hydrodynamics}

The most general ideal Lagrangian corresponds to terms where $T_i^\mu$ and $J^\mu$ contain as few derivatives as possible, as hydrodynamics is a gradient expansion in derivatives. We have
\begin{align}\label{eq:ideal L app}
    \cL_{\mathrm{ideal}} = \left(p\delta^j_i +\pi_i u^j+ \kappa^j_k \p_i\sigma^k \right)E^i_{a,j} + mnu^i E'_{a,0i} + n C_{a,0},
\end{align}
with $\pi^i+\kappa^0_j \p_i \sigma^j = m n u^i$, which comes from \eqref{eq:momentum current identity}.
Under the KMS transformation, we find,
\begin{align}
    &-\ii T_0(\tilde \cL_{\mathrm{ideal}} - \cL_{\mathrm{ideal}}) \nonumber\\
    &= p \delta^\mu_i \p_\mu u^i + \pi_i u^\mu \p_\mu u^i + \kappa^\mu_j \p_i \sigma^j \p_\mu u^i +n u^\mu \p_\mu \mu \nonumber\\
    &= \p_\mu(p u^\mu) - u^\mu \p_\mu p +\pi_i u^\mu \p_\mu u^i +\p_\mu\left(\kappa^\mu_j u^\nu \p_\nu \sigma^j\right) -\p_\mu\kappa^\mu_j u^\nu \p_\nu \sigma^j - \kappa^\mu_j u^\nu \p_\nu \p_\mu \sigma^j   + n u^\mu \p_\mu \mu \nonumber\\
    & = \p_\mu(p u^\mu) - u^\mu \p_\mu p +\pi_i u^\mu \p_\mu u^i- \kappa^\mu_j u^\nu \p_\nu \p_\mu \sigma^j   + n u^\mu \p_\mu \mu,
\end{align}
where in the last step we used $\beta_0 u^\mu \p_\mu \sigma^i(x)  = \delta^i_I\ud\sigma^I/\ud \sigma^t =0$. Now, the change is a total derivative provided we demand the equation of state
\begin{align}
    \delta p = \pi_i \delta u^i - \kappa^\mu_i \delta(\p_\mu \sigma^i) + n\delta \mu.
\end{align}
By comparing \eqref{eq:ideal L app} with \eqref{eq: eff L}, we find the ideal stress tensor and current as
\begin{subequations}
    \begin{align}
        T^\mu_{i,(0)} & = p \delta^\mu_i +\pi_i u^\mu + \kappa^\mu_j \p_i \sigma^j,\\
        J^\mu_{(0)} &= n u^\mu,
    \end{align}
\end{subequations}
and they satisfy \eqref{eq:momentum current identity}.

\subsection{Next-to-leading-order hydrodynamics}

The dissipative Lagrangian involving $E'_{a,0i}$ and $C_{a,0}$ can be removed by field redefinitions \cite{Crossley:2015evo}. Suppose we have the first order dissipative Lagrangian as
\begin{align}\label{eq:L1 redef}
    \cL^{(1)} \sim - A^i E'_{a,0i} - B C_{a,0}.
\end{align}
Consider a shift of the $r$-fields, $\mu\to \mu+\delta \mu$ and $u^i\to u^i +\delta u^i$, in the ideal Lagrangian $\delta \cL_{\mathrm{ideal}} = \delta n_0 C_{a,0} + \delta a^i E'_{a,0i}$ with $\delta n_0 = \chi \delta \mu $ and $\delta a^i = m \delta n_0 u^i + m n_0 \delta u^i$. Then, by choosing $\delta n_0 = B$ and $\delta a^i = A^i$, \eqref{eq:L1 redef} can be removed, provided we ignore its contribution to the bulk viscosity.


The most general KMS-invariant Lagrangian is given by
\begin{align}\label{eq:next L}
    \cL^{\prime} =  \ii T_0 s^{ijkl}U_{a,ij}\left(U_{a,kl} + \ii \beta_0 \p_k \p_l \mu\right)+\ii T_0 t^{ijkl}E_{a,ij}\left(E_{a,kl} + \ii \beta_0 \p_k u_l\right)  +t^{ijkl}_{\mathrm{odd}} E_{a,ij} \p_k u_l,
\end{align}
where $t^{ijkl}_{\mathrm{odd}} = t_{\mathrm{odd}}(\epsilon^{ik}\delta^{jl} + \delta^{ik}\epsilon^{jl})$.   The last term proportional to $t_{\mathrm{odd}}$ is $\Theta$-even and corresponds to non-dissipative dynamics---it describes the Hall viscosity in fluid dynamics. Upon variation, the equations of motion become
\begin{align}\label{eq:eom full}
    B n_0 \epsilon_{ij}\p_0\phi^j  + \chi^{-1}n_0\p_i \delta n + a^{ijkl}\p_j\p_k \phi_l+t^{ijkl}\p_j\p_k\p_0 \phi_l - t^{ijkl}_{\mathrm{odd}}\p_j\p_k\p_0 \phi_l&=0, \nonumber\\
    \p_0 \delta n - n_0 \p_0 \p_i \phi^i + \chi^{-1}(s_1+s_2)\p^4 \delta n  &= 0.
\end{align}
The $t_{\mathrm{odd}}$-term can be effectively absorbed into the Lorentz force term by shifting $B \to B' \equiv B  - n_0^{-1}t_{\mathrm{odd}}k^2$. Upon Fourier transformation, we find the normal modes as
\begin{subequations}
    \begin{align}
        \omega &= \pm c k^2 \pm c'k^4 - \ii \Gamma k^4,\\
        \omega &= - \ii D k^4,
    \end{align}
\end{subequations}
where
\begin{align}
    c &= \frac{\sqrt{a_2(a_1+a_2+\chi^{-1}n_0^2)}}{B n_0}, \nonumber\\
    c'& = \frac{t_{\mathrm{odd}}c}{B n_0},\nonumber\\
    D &= \frac{(a_1+a_2)(s_1+s_2)}{\chi(a_1+a_2)+n_0^2},\nonumber\\
    \Gamma & =\frac{D n_0^2}{2\chi(a_1+a_2)} + \frac{a_2 t_1+(a_1+2a_2)t_2}{2 (B n_0)^2} .
\end{align}
We see the term coming from Hall viscosity contributes as higher-order corrections---the same order as the dissipation---to the propagating mode.

\subsection{Long-range Coulomb interactions}
We briefly discuss how long-range Coulomb interactions modify the EFT above.  Density-dependent interactions with a potential $1/r^{2-\alpha}$ effectively cause the bulk modulus and charge susceptibility to scale as $a_1\sim \chi^{-1}\sim k^{-\alpha}$ \cite{Lee_Pinning}. In the linear response regime, we find that the magnetophonon disperses as $\omega \sim \pm k^{2 - \alpha/2} - \ii k^{4-\alpha}$, and the charge relaxes as $\omega\sim -\ii k^{4-\alpha}$. The scaling analysis indicates that $\delta n \sim k^{(d+\alpha)/2},\;\psi^i\sim k^{(d+\alpha)/2-1}$, and it implies the nonlinear terms $\lambda^{ijkl}\sim \lambda'\sim k^{(4-3\alpha-d)/2}$ are irrelevant when $d>4-3\alpha$. For unscreened Coulomb potential $\alpha=1$, we generalize the dispersion relation in \cite{Bonsall_some} to dissipative regime with $\omega \sim \pm k^{3/2}-\ii k^3$, which is stable at linear order. For $0<\alpha<2/3$, we find the linear response theory at $d=2$ is unstable.  

Lastly, $\chi_{\phi^i \phi^i}\sim 1/k^{2-\alpha}$, meaning that the Mermin-Wagner Theorem does not preclude long-range order in $\langle \phi^i \phi^j\rangle$ for any $\alpha>0$.  When $\alpha=0$, we expect the EFT to break down on extremely large length scales where the quantum Hall crystal will inevitably melt at finite temperature, exhibiting only quasi-long-range order.

\section{Galilean symmetry in the Hamiltonian model}\label{app:galileanH}
In the main text, we described a Hamiltonian model for particles in the lowest Landau level in a quantum Hall crystal.  As a reminder, we considered the displacement $q_i(\mathbf{n})$ of particles at integer-labeled lattice site $\mathbf{n}$;  let $\bar x_i(\mathbf{n})$ be the equilibrium spatial coordinate of each such particle.  We have a Lagrangian of the form 
\begin{align}\label{eq:LappB}
    L = \sum_{\mathbf{n}}\left[\frac{B e_{\mathbf{n}}}{2} \epsilon_{ij}  q_i(\mathbf{n}) \dot q_j(\mathbf{n}) - H(\Delta_i q_i(\mathbf{n})) + e_{\mathbf{n}}E_i q_i(\mathbf{n})\right],
\end{align}
where $E_i$ is a constant external electric field, and we have added the magnetic field $B$ explicitly above.  The Euler-Lagrange equations of \eqref{eq:LappB} encode Hamilton's equations, including the background electric field.  In the model of the main text, we had set $e_{\mathbf{n}}=1$, which we interpret as all charged particles having the same charge.  It will be interesting for the discussion here to allow it to be arbitrary.

Under a Galilean boost transformation, \begin{equation}
    q_i(\mathbf{n}) \rightarrow q_i(\mathbf{n}) + v_i t \label{eq:appendixqshift}
\end{equation} where $v_i$ is a constant for all particles at distinct lattice sites $\mathbf{n}$; due to the presence of a magnetic field, in the non-relativistic limit, we must also transform \begin{equation}
    E_i \rightarrow E_i - \epsilon_{ij}v_j B, \label{eq:appendixEshift}
\end{equation}
which we remind the reader is a consequence of the boost invariance of the linear combination $E_i + B \epsilon_{ij} \dot q_j$.   One readily checks that $L$ is invariant under these transformations. 

Because of this boost symmetry, we can convert any microscopic solution of the Euler-Lagrange equations $\bar q_i(\mathbf{n},t)$ into another solution $\bar q_i(\mathbf{n},t) + c_i t$ at the cost of simply adjusting the electric field according to \eqref{eq:appendixEshift}.  In particular, consider the expectation value of the current operator \begin{equation}
    J_i(x) = \sum_{\mathbf{n}} \delta(q_i(\mathbf{n}) + \bar x_i(\mathbf{n})-x_i) \dot q_i(\mathbf{n}) e_{\mathbf{n}}
\end{equation}
in a state with zero electric field:  assuming that we have not completely broken rotational symmetry we must have $\langle J_i\rangle=0$ in a state with $E_i=0$.  The boost symmetry allows us to take an ensemble of solutions to Hamilton's equations with various initial conditions and transform them to a new ensemble of solutions in a finite electric field by simply applying \eqref{eq:appendixqshift} and \eqref{eq:appendixEshift}.  Assuming that a typical equilibrium state has statistical homogeneity, we see that  \begin{equation}
    \langle J_i(x)\rangle_{E_i \ne 0} = \langle J_i(x)\rangle_{E_i=0} + \frac{1}{B}\epsilon_{ij}E_j \left\langle \sum_{\mathbf{n}} \delta(q_i(\mathbf{n}) + \bar x_i(\mathbf{n})-x_i)e_{\mathbf{n}} \right\rangle = \frac{n_0}{B}\epsilon_{ij}E_j,
\end{equation}
where $n_0$ is the net charge density of the theory.  (We assume that $e_{\mathbf{n}}$ varies rapidly enough in space that any coarse-grained density operator has uniform expectation value.) 
This implies that the longitudinal conductivity $\sigma_{xx}=0$ and $\sigma_{xy}=n_0/B$, in agreement with the result derived in the main text.  


We do caution the reader that the vanishing of $\sigma_{xx}$ \emph{by itself} is not enough to rule out the presence of incoherent conductivity in the absence of an intrinsic momentum relaxation mechanism \cite{Hartnoll:2007ih},  which our Hamiltonian model does not have.  Of course, as we have seen, the covariance of $L$ under boosts ensures that there is no incoherent conductivity.

We can physically interpret this Hamiltonian model as describing a limit where all charge is ``fixed charge" in the terminology of \cite{Halperin_Hallcrystal}.   In such models in the lowest Landau level regime, we evidently always have a Galilean symmetry arise.  To break this symmetry, one must consider the fluctuations of thermally mobile quasiparticles in addition to the dynamics of the quantum Hall crystal.  If there is a gap $\Delta$ to such freely excited quasiparticles, we would then expect that $\sigma_0 \sim \exp[-\Delta/T]$, implying a parametrically long time scale over which the $z\approx 3$ universality class of dynamics observed in the main text would be detectable in experiment.
\end{appendix}

\end{document}